# Single-shot thermal ghost imaging using wavelength-division multiplexing

Chao Deng,[1,2] Jinli Suo,[1] Yuwang Wang,[1] Zhili Zhang,[2] and Qionghai Dai[1,a)]
[1]*Department of Automation, Tsinghua University, Beijing 100084, China*
[2]*High-Tech Institute of Xi'an, Xi'an 710025, China*

Ghost imaging (GI) is an emerging technique that reconstructs the target scene from its correlated measurements with a sequence of patterns. Restricted by the multi-shot principle, GI usually requires long acquisition time and is limited in observation of dynamic scenes. To handle this problem, this paper proposes a single-shot thermal ghost imaging scheme via a wavelength-division multiplexing technique. Specifically, we generate thousands of correlated patterns simultaneously by modulating a broadband light source with a wavelength dependent diffuser. These patterns carry the scene's spatial information and then the correlated photons are coupled into a spectrometer for the final reconstruction. This technique increases the speed of ghost imaging and promotes the applications in dynamic ghost imaging with high scalability and compatibility. © 2018 Author(s). All article content, except where otherwise noted, is licensed under a Creative Commons Attribution (CC BY) license (http://creativecommons.org/licenses/by/4.0/).
https://doi.org/10.1063/1.5001750

As an emerging imaging technique, ghost imaging (GI) has attracted extensive attention in recent years due to its advantages in some harsh imaging environments. Ghost imaging can be classified, in terms of light sources, into three categories: quantum GI,[1,2] classical GI,[3–6] and computational GI.[7,8] Benefiting from its great potential, GI has been applied in various fields including remote sensing,[9,10] optical encryption,[11,12] scattering robust imaging,[13,14] and object tracking.[15]

Unlike conventional imaging methods, a large number of measurements need to be performed to obtain an image, which essentially reduces the imaging speed. The number of required measurements is roughly determined by the target pixel resolution. To form a $64 \times 64$-pixel image, for instance, we need approximately 4096 illumination patterns employing the correlation algorithm[16–18] and at least 1024 via the compressed sensing (CS) algorithm.[19–21] These illumination patterns form a sampling matrix of the target scene and are required to be uncorrelated with each other in order to guarantee the matrix's reversibility (for the correlation algorithm) or satisfy restricted isometry property[22] condition (for the CS algorithm). As a consequence, the speed of a ghost image setup is limited by the refresh rate of uncorrelated speckles (thermal GI) or uncorrelated programmed patterns (computational GI), so ghost imaging is excluded from practical imaging systems for dynamic scenes. Besides limitation in capturing moving scenes, the long acquisition time of GI also suffers from imaging quality degeneration caused by illumination fluctuation, background noise, and detector instability during the sequential sampling process.

To speed up the imaging process of GI, a straightforward solution is to generate uncorrelated patterns fast through various strategies. However, fast generation of sufficient uncorrelated patterns for reconstruction using conventional reconstruction algorithms is really challenging. Since introduced into GI,[18,23] the alternating projection (AP) algorithm has attracted a lot of attention.[24–30] The AP algorithm imposes no strict limitation on the un-correlation among illumination patterns. Mathematically, AP is widely used for solving a linear equation system by alternatively projecting to constraint/subspaces of different equations. This algorithm does not impose constraint on the correlation among different linear equations. As for GI, each correlated measurement and corresponding illumination pattern carry some scene information and act as a support constraint of the latent solution in the Fourier and spatial domains. The measurement of each newly added correlated pattern can still carry a little extra information of the object, which can be accumulated by AP to infer the target object, even in the cases of strongly correlated patterns. In other words, the AP algorithm decreases the sensitivity of the GI scheme to the presence of uncorrelated patterns. Therefore, in this paper, we resort to producing correlated speckle patterns fast and adopting the AP algorithm for reconstruction. Inspired by the differentiation of speckle patterns when dissimilar wavebands pass through a thin diffuser, we propose a scheme that produces required patterns simultaneously rather than sequentially using the wavelength-division multiplexing technique. In the sense that multiple samplings are fulfilled at the same time, we call this "single-shot."

Similar to conventional thermal GI, we exploit a thin diffuser to yield patterned illumination. Speckles emerge when a narrowband laser penetrates a scattering medium, which is extremely sensitive to the wavelength.[31] The surface property of the diffuser and coherence and polarization of incident light can also influence the detailed speckle intensity distribution.[31] Some researchers have already reported the optical performance of thin diffusers and exploited it in imaging systems including GI. With a femtosecond laser of controlled bandwidth, Curry *et al.* presented the diffusion properties of a strongly scattering material by measuring the speckle contrast, which showed that speckle contrast decreased with the increasing light source bandwidth and diffuser thickness.[32] Andreoli *et al.* measured the spectrally

a)qhdai@tsinghua.edu.cn

resolved transmission matrix (TM) of a strongly scattering medium using a spatial light modulator (SLM) and discussed the spectral correlation function of the medium.[33] Based on aforementioned work, Kolenderska et al. performed a single-shot 2D incoherent imaging setup through scattering medium using spatial-spectral encoding.[34] Shin et al. utilized a $TiO_2$-tipped single mode fiber (SMF) as a thin diffuser by the wavelength-adjustable laser source to sequentially generate uncorrelated speckles for GI,[35] and they could not achieve single-shot ghost imaging. Nandan proposed a spectral ensemble ghost imaging technique and implemented simulations for a 1D system within one shot.[36]

Differently, we display a large amount of speckle patterns simultaneously to perform single-shot ghost imaging. In detail, the process of a monochromatic coherent beam passing through a random scattering medium, although complex, remains linear and the corresponding speckle emerges. During light transmission, the random scattering medium responses differently to each spectral correlation bandwidth $\Delta\lambda = \sqrt{\lambda^2/(c\Delta t)}$ (where $\lambda$ is the wavelength, c is the speed of light, and $\Delta t$ is the coherence time of the light source). Therefore, we can consider the outgoing broadband light beam as the intensity summation of speckles with varying spectrums and spatial distributions. In other words, the spectrum varies at different positions. Using a spectrometer instead of a bucket detector, we can collect a large number of correlated measurements simultaneously and achieve fast ghost imaging.

The large number of patterns corresponding to different spectrums is not strictly uncorrelated. Ideally, by increasing the spectrum range of the light source and decreasing the spectral correlation bandwidth using strongly scattering medium (e.g., layer of ZnO or $TiO_2$ powder of specified thickness), we can generate sufficient uncorrelated speckle patterns for better reconstruction. However, the intensity summation of uncorrelated speckles would reduce the contrast of the speckle field[37] and make it difficult to locate the speckle plane. In implementation, we trade off between the speckle contrast and the number of uncorrelated patterns for better experimental results.

Figure 1 depicts our single-shot thermal ghost imaging setup. We utilize a tungsten lamp house (HILT250: 300 nm–2500 nm) as our light source. The incoming white light enters a diffuser (holographic diffuser: 54506, Edmund) to produce the speckle field and then arrives at the object plane after magnification by a convex lens 1 (L1). The outgoing light is coupled by a convex lens 2 (L2) into a point spectrometer (CCS200/M, Thorlabs: 200 nm–1000 nm, resolution: 0.22 nm). We replace the bucket detector with a spectrometer for collecting the measurements corresponding to a large number of patterns of different wavelengths simultaneously. As shown in the upright inset of Fig. 1, the speckle field after the holographic diffuser is composed of speckles corresponding to different wavelengths. Theoretically, the spectral correlation bandwidth of a scattering medium is inversely proportional to the square of its thickness.[38] The holographic diffuser in our experiment is coated with a thin film about 5 μm thick, and one can calculate that the spectral correlation bandwidth is approximately 20 nm. It is worth mentioning that our technique is only applicable to gray objects rather than colorful ones, as a broadband light source is required for wavelength-division multiplexing. However, gray objects widely exist in natural scenes, especially in the micro-regime, so the proposed approach is not general but still of wide applicability.

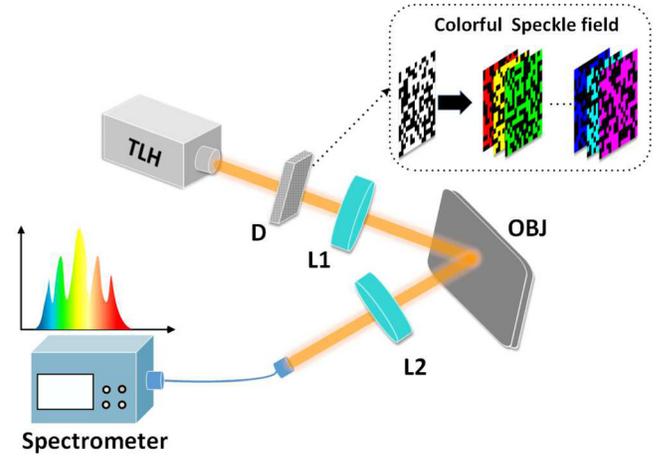

FIG. 1. Experimental setup of the single-shot thermal ghost imaging. TLH, tungsten lamp house (300 nm–2500 nm); D, holographic diffuser; L1 and L2 convex lenses; OBJ, object. A white light source enters a holographic diffuser to generate the colourful speckle field. As the thickness of the holographic diffuser is around 5 μm, the spectral correlation bandwidth is approximately 20 nm. The point spectrometer is used to calibrate the speckle field on the object plane and functionally equivalent to a bucket detector.

Denoting the spectrum at the speckle plane and object as $I(x, \lambda)$ and $O(x)$, the measurements of the spectrometer $S(\lambda)$ can be written as

$$S(\lambda) = \int I(x, \lambda)O(x)dx, \quad (1)$$

where x and $\lambda$ denote the 2D coordinate on the object plane and the wavelength, respectively. Because different speckle patterns corresponding to different $\lambda$ are displayed simultaneously, we can implement a large number of measurements at one time so as to achieve singe-shot ghost imaging.

To reconstruct the object, we need to calibrate the set of speckle patterns $\{I(x, \lambda)\}$ first and then perform image reconstruction by solving the following minimization:

$$\begin{aligned} \min_{\mathbf{O}} \quad & \|\mathbf{S} - \mathbf{I} \cdot \mathbf{O}\|_2 \\ \text{s.t.} \quad & \mathbf{O} \geq 0, \end{aligned} \quad (2)$$

where $\mathbf{O}$ is the vectorized form of $O(x)$, $\mathbf{S}$ is a vector denoting the measurement of the spectrometer, and $\mathbf{I}$ is the sampling matrix with each row representing each speckle pattern of a specific wavelength.

In terms of reconstruction, the AP algorithm essentially reduces the fitting error by incorporating the constraint from the patterns $\mathbf{I}$ and correlated measurements $\mathbf{S}$ in spatial and Fourier domains alternatively. The reconstruction process includes three main steps

First, we denote $\tilde{I}(x, \lambda_n) = I(x, \lambda_n)O(x)$ and have

$$\mathcal{F}\{\tilde{I}(x, \lambda_n)\}(0) = \int_x \tilde{I}(x, \lambda_n) \exp(-j2\pi kx)dx|_{k=0}$$
$$= \int_x \tilde{I}(x, \lambda_n)dx = S(\lambda_n), \quad (3)$$

where $\mathcal{F}\{\cdot\}$ denotes the Fourier transform, and $n$ indexes the pattern number ($n = 1, 2, 3...$). Based on this, with the initial guess of the object $O(x)$, we utilize the measurement $S(\lambda_n)$ to update $\tilde{I}(x, \lambda_n)$ in the Fourier domain as

$$\mathcal{F}\{\tilde{I}^{update}(x, \lambda_n)\} = \mathcal{F}\{\tilde{I}(x, \lambda_n)\} \cdot (1 - \delta(x)) + S(\lambda_n) \cdot \delta(x). \quad (4)$$

Here, $\delta(x)$ is the discrete delta function.

Second, we update the object $O(x)$ in the spatial domain using the following equation:

$$O^{update}(x) = O(x) + \frac{I(x, \lambda_n)}{\max\{I^2(x, \lambda_n)\}} \cdot \left[\tilde{I}^{update}(x, \lambda_n) - O(x)I(x, \lambda_n)\right]. \quad (5)$$

Third, we repeat the above two steps for all the measurements until the algorithm converges. The number of loops is dependent on the target resolution, and in our experiments, it is typically $100 \sim 200$ for $16 \times 16$ pixel imaging.

Similar to classical GI and computational GI, we use the set of 2D patterns $\{I(x, \lambda)\}$ to resolve the spatial information of the target scene, and these patterns are requisite input of the reconstruction algorithm. The difference is that classical GI uses a high resolution CCD to record the patterns, computational GI uses the computer calculated patterns, while we calibrate them beforehand. Specifically, we use a digital micromirror device (DMD, Texas Instruments Discovery 4100) to traverse each spatial position $x$ of $\{I(x, \lambda)\}$ and the spectrometer to measure the corresponding spectrum. Considering the spatial correlation of neighbouring pixels and the optical intensity of the scanning pattern, the image macro-pixel size is the same as the speckle grain size, which is about $41 \mu m \times 41 \mu m$. As the DMD pixel size is $13.68 \mu m \times 13.68 \mu m$, we group DMD elements into $3 \times 3$ super-pixels, to match the spatial resolution of speckles. Afterwards, we figure out the set of speckle patterns $\{I(x, \lambda)\}$ on the speckle plane. Limited by the wavelength range of the spectrometer (200 nm–1000 nm), together with the filtering effect of diffusers and lenses, the ultimate wavelength range is 391 nm $\sim$850 nm with approximately 2000 patterns in total. The imaging resolution is dependent on the information coding ability of these patterns. Empirically, we can implement image reconstruction of around 250 pixels, so we set our imaging resolution as $16 \times 16$. The correlation of the speckle field is shown in Fig. 2, and the correlation between two speckle patterns $I(x, \lambda_1)$ and $I(x, \lambda_2)$ is calculated according to the following equation:

$$corr = \frac{\sum \left[I(x, \lambda_1) - \overline{I(x, \lambda_1)}\right]\left[I(x, \lambda_2) - \overline{I(x, \lambda_2)}\right]}{\sqrt{\sum \left[I(x, \lambda) - \overline{I(x, \lambda_1)}\right]^2 \sum \left[I(x, \lambda_2) - \overline{I(x, \lambda_2)}\right]^2}}. \quad (6)$$

We can see that these speckle patterns are not strictly uncorrelated with others. As the spectral correlation bandwidth of the holographic diffuser is approximately 20 nm, the total number of uncorrelated patterns is just about $(850 - 391)/20 \approx 23$. We calculate the rank of the calibrated sampling matrix **I** (i.e., 26), which verifies that the experiment is consistent with the theoretical derivation.

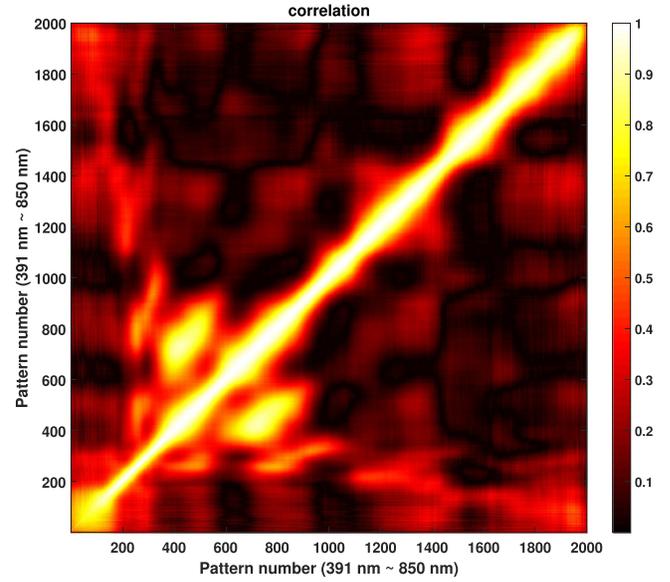

FIG. 2. Pattern correlation of the speckle field at different wavelengths in our experiment. After the filtering effect of all optical elements, the optical wavelength ranges from 391 nm to 850 nm, with about 2000 different speckle patterns in total and about 23 uncorrelated speckle patterns.

To demonstrate the imaging performance of our approach, we use our setup to reconstruct several static images, using the digits "3," "6," and "9" from the MNIST Database as our test scenes. Figure 3 shows the result, together with the correlated measurements. The first and second rows display the ground truth and reconstruction of three digits separately. The bottom row shows the corresponding spectrum. From the result, we can see that although some undesired artifacts exist, the imaging quality is rather high with a satisfactory signal-to-noise ratio (SNR). The slight artifacts are mainly attributed to the correlation among speckle patterns, and some specially fabricated diffuser may reduce the correlation and improve the final reconstruction.

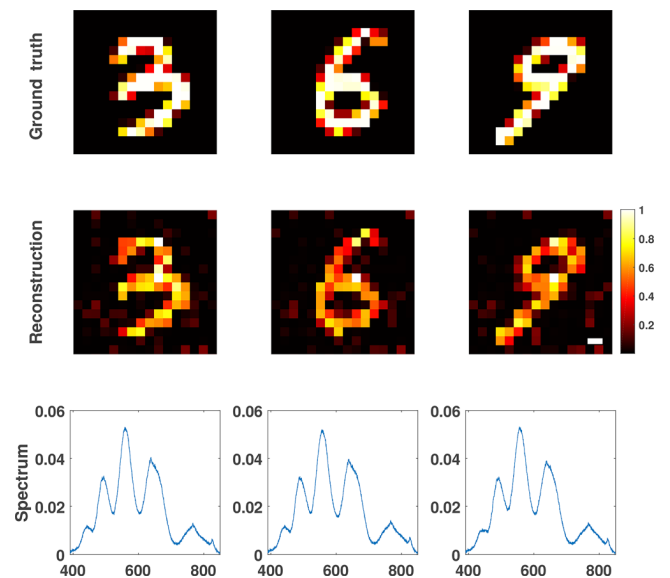

FIG. 3. Reconstruction of handwritten digits 3, 6, and 9 from the MNIST Database. The actual image of digits is in the first row; reconstruction from our experiment is in the middle row. The corresponding spectrum is in the third row. All the colorbars are the same in the first two rows and just shown in the last panel of the second row. Scale bar: 80 $\mu m$.

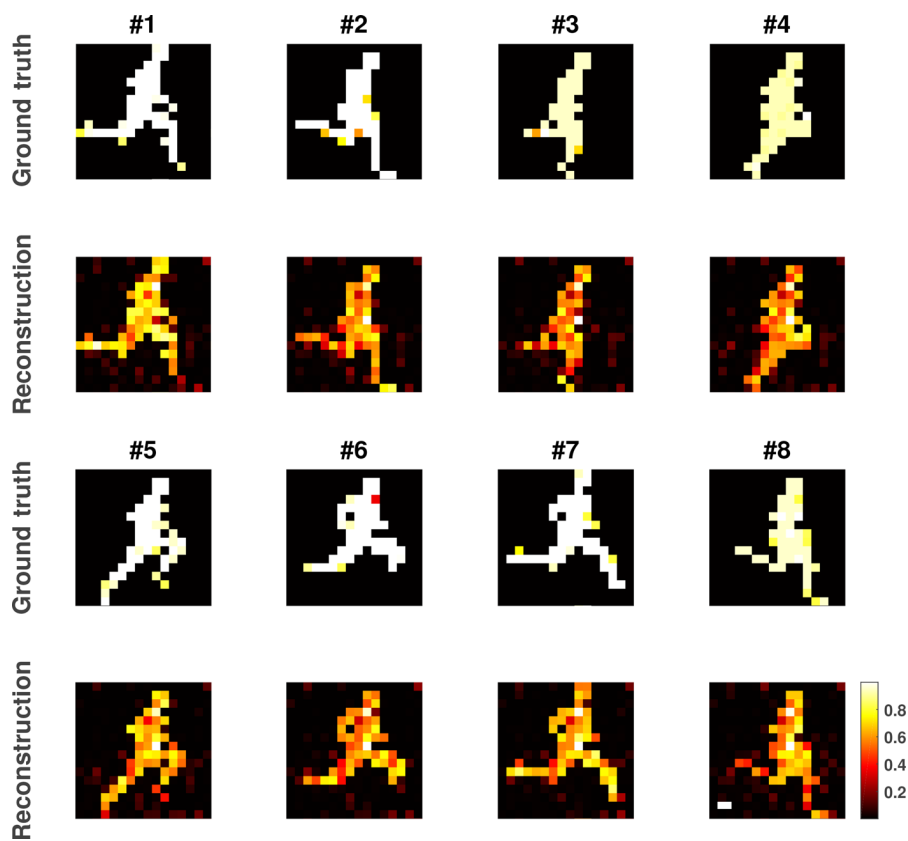

FIG. 4. Reconstruction of a dynamic scene. Ground truth images of eight frames are in the first and third rows. The corresponding reconstruction is in the second and fourth rows. All the colorbars are the same as shown in the last panel. Scale bar: 80 $\mu m$.

We also conduct an experiment to validate the dynamic imaging capability of our single-shot ghost imaging setup. We display a "running-man" animation of eight frames as our sample, and the reconstruction result is shown in Fig. 4. The first and third rows show eight frames of ground-truth images, and the corresponding reconstruction is in the second and fourth rows. We set the DMD and spectrometer synchronously controlled by the computer to collect the spectrum of each frame. The frame rate of our setup is only restricted by the integration time of the spectrometer, which is about 50 ms in our experiment. High speed dynamic imaging can be fulfilled with a faster spectrometer.

Here, we specially compare our method with SLM-based ghost imaging using the CS algorithm for reconstruction, which can also achieve dynamic imaging with the DMD working at full frame rate.[39] First, our setup is advantageous in scalability and compatibility. With a high-end spectrometer, the pixel resolution and imaging speed can both be increased significantly, while the SLM-based setup is largely limited by the SLM's performance. The light modulation part can be compatible with most off-the-shelf spectrometers, and one can choose the specifications according to the application scenarios. Second, the proposed setup can work for imaging beyond the visible light spectrum by a corresponding spectrometer, but the SLM-based counterpart might be limited due to lacking available SLMs.

In conclusion, we design a single-shot thermal ghost imaging setup using wavelength-division multiplexing and we validate its effectiveness. A holographic diffuser is used to decompose the broadband light source into thousands of speckle patterns to achieve single-shot ghost imaging. We implement image reconstruction with correlated speckle patterns using the alternating projection (AP) algorithm.

Our technique speeds up the process of ghost imaging and promotes its practical applications. Benefiting from its fast acquisition, the proposed approach can track dynamic scenes. There are two limitations in our experiment. First, a broadband light source is indispensable and thus our system is mainly limited to gray scale scenes. However, considering that gray objects/samples are rather common in both macro- and micro-regimes, the approach is still widely applicable. Second, the pixel resolution is only limited by the spectrometer. With the high-end spectrometer (e.g., OSA201C, Thorlabs: 350 nm–1100 nm, resolution: 5 pm), we can improve the pixel resolution significantly. While tracking dynamic scenes, we can exploit temporary redundancy among video frames to decrease the required patterns for further resolution enhancement. In addition, our approach can be readily realized in the infrared, X-ray, and terahertz fields. We expect that this technique can promote the development of ghost imaging.

The authors acknowledge the financial support by the National Natural Science Foundation of China (NSFC) (Nos. 61327902 and 61627804).